\documentclass[sigconf,screen,nonacm]{acmart}
\renewcommand\footnotetextcopyrightpermission[1]{}
\usepackage{graphicx} 
\usepackage{listings}
\usepackage{multirow}
\usepackage{amsmath}
\usepackage{stmaryrd}
\usepackage{mdframed}
\usepackage{tcolorbox}
\usepackage{multicol}
\usepackage{caption}
\usepackage{xspace}
\usepackage{xurl}
\usepackage{amsthm}
\usepackage{dirtree}
\usepackage{todonotes}
\usepackage[bottom]{footmisc}
\usepackage[inline]{enumitem}
\usepackage{adjustbox}
\usepackage{booktabs}
\usepackage{placeins}
\usepackage{forest}
\lstdefinestyle{mystyle}{
	basicstyle=\ttfamily\scriptsize,
	breakatwhitespace=false,         
	breaklines=true,                 
	captionpos=b,                    
	keepspaces=true,                 
	numbers=left,                    
	numbersep=5pt,                  
	showspaces=false,                
	showstringspaces=false,
	showtabs=false,                
	tabsize=2,
	frame=single,
	xleftmargin=3.0ex
}

\newcommand{\bfs}{\textit{obfs}\xspace}
\newcommand{\gaoss}{\textit{gaoss}\xspace}

\newcommand{\macaron}{\textit{macaron}\xspace}
\newcommand{\ossr}{\textit{oss-rebuild}\xspace}
\newcommand{\Macaron}{\textit{Macaron}\xspace}
\newcommand{\Ossr}{\textit{Oss-rebuild}\xspace}

\newcommand{\pypi}{\textit{PyPI}\xspace}

\newcommand{\daleq}{\textit{daleq}\xspace}
\newcommand{\daleqpy}{\textit{daleq4py}\xspace}
\NewDocumentCommand{\filepath}{v}{\path{#1}}

\newtcolorbox{rqanswer}[1]{
	colback=black!4,
	colframe=black!60,
	fonttitle=\bfseries,
	title={#1},
	boxrule=0.6pt,
	arc=1.5pt,
	left=4pt, right=4pt, top=3pt, bottom=3pt
}

\title{No Snake Oil: Verifying Python Package Builds}

\author{Jens Dietrich}
\email{jens.dietrich@vuw.ac.nz}
\affiliation{%
	\institution{Victoria University of Wellington}
	\city{Wellington}
	\country{New Zealand}
}
\author{Spencer Sun}
\email{spencer.sun@vuw.ac.nz}
\affiliation{%
	\institution{Victoria University of Wellington}
	\city{Wellington}
	\country{New Zealand}
}
\author{Tim W. White}
\email{walton.white@oracle.com}
\affiliation{%
	\institution{Oracle Inc}
	\city{Wellington}
	\country{New Zealand}
}
\author{Behnaz Hassanshahi}
\authornotemark[1]
\email{behnaz.hassanshahi@oracle.com}
\affiliation{%
	\institution{Oracle Inc}
	\city{Brisbane}
	\country{Australia}
}

\begin{abstract}
	Python has become the default language for interacting with AI, with packages being distributed through registries like the Python Package Index (\pypi). This creates a need to analyse supply chains comprising such packages.
	One such analysis is to rebuild packages in order to identify compromised builds injecting malware.
	Independent rebuilds in hardened environments have the added advantage that they can generate and record provenance in order to increase the trustworthiness of packages.
	Two tools that are designed to automate such rebuilds and run them at scale are \macaron and \ossr. We study 12,180 popular releases from \pypi and find that the byte-for-byte equivalence
	rate is generally low. We analyse the reasons why they produce different wheels, and find that equivalence between the original and rebuilt wheels can often still be established, preserving most of the guarantees users expect from rebuildable releases.
	We present and evaluate \textit{daleq4py}, a tool to establish the equivalence of Python wheels through the kernel of a normalisation function that is based on provenance-preserving datalog rules. 
	Experimental results show that \daleqpy substantially expands the set of rebuilds that can be accepted as equivalent.
	Although only 15.4\% of \macaron rebuilds and 19.1\% of \ossr rebuilds are byte-for-byte identical to the published \pypi wheels, \daleqpy establishes wheel equivalence for 60.2\% and 78.9\% of source-equivalent rebuilds, respectively.
	 
\end{abstract}

\begin{document}
\maketitle

\section{Introduction}
\label{sec:introduction}

Modern software is built on top of a deep stack of open-source dependencies, distributed through package registries.
While this enables economies of scale and rapid innovation, it also creates a significant attack surface.
A series of high-profile incidents -- \textit{solarwinds}, \textit{codecov}, \textit{equifax}, \textit{log4shell}~\cite{ellison2010evaluating,martinez2021software,enck2022top}, the \textit{xz} backdoor~\cite{xz}, and more recently \textit{react2shell}~\cite{react2shell} and \textit{shai-hulud}~\cite{ShaiHulud} -- have brought software supply chain security to the top of the policy agenda~\cite{EO14028}.
A particular type of attack targets the automated processes used to build and distribute packages, injecting backdoors and other malware. 
The possibility of such attacks was already anticipated in Ken Thompson's Turing award speech~\cite{thompson1984reflections}, and a broad body of work catalogues their growing prevalence~\cite{ohm2020backstabber,andreoli2023prevalence}.

A widely discussed countermeasure is to reproduce builds~\cite{reproduciblebuilds,lamb2021reproducible}.
The basic idea is straightforward: a build is repeated in an independent, hardened environment, and the resulting binary is compared to the published artifact, typically by means of cryptographic hashes.
An attacker who has compromised the original build pipeline can no longer hide, since the independent build will produce a different binary that can easily be detected~\cite{enck2022top,butler2023business,fourne2023s}.
Two builds starting from the same sources should produce the same package:

\vspace{-0.3cm}
\begin{equation}
src_1 = src_2 \Rightarrow build_1(src_1) = build_2(src_2)
\label{eq:repro}
\end{equation}

When such rebuilds are run at scale, they also generate independent provenance and increase the level of trust users can place in packages~\cite{slsa,source-prov}.
In practice, reproducing builds is challenging and often fails~\cite{shi2021experience,drexel2024reproducible,randrianaina2024options,bajaj2024unreproducible,malka2024reproducibility}: builds are sensitive to compilers and compiler or build plugin versions, file orderings, timestamps, environment variables, and a long tail of other inputs that are rarely captured by package metadata.
Two complementary lines of work address this.
The first attempts to lock down the build environment sufficiently to make builds deterministic.
The \textit{Reproducible Builds} project~\cite{reproduciblebuilds} provides cross-ecosystem guidance and tooling for eliminating common sources of non-determinism, such as embedded timestamps, build paths and non-deterministic file orderings.
Several ecosystems complement this with \emph{lock files} that pin direct and transitive dependencies -- and increasingly the build plugins themselves -- to specific versions, so that the inputs to a build are fully determined~\cite{schmid2025maven}.
The second line accepts that bit-for-bit reproduction is often unachievable and instead diffs build outputs to support security assessment, using tools such as \textit{diffoscope}~\cite{diffoscope}, \textit{BuildDiff}~\cite{macho2017extracting}, \textit{RepLoc}~\cite{ren2018automated} and \textit{RepTrace}~\cite{ren2019root}.

A semi-formal middle ground is to weaken the consequent of \eqref{eq:repro} from binary equality to an \emph{explainable equivalence} relation $\simeq_{bin}$ that under-approximates behavioural equivalence and tolerates benign variability such as differences in compilers, file orderings, or timestamps~\cite{dietrich2025levels,dietrich2025daleq}:

\vspace{-0.3cm}
\begin{equation}
src_1 = src_2 \Rightarrow build_1(src_1)\simeq_{bin}build_2(src_2)
\label{eq:equiv}
\end{equation}

The \textit{daleq} tool~\cite{dietrich2025daleq} implements such a relation for Java bytecode and was shown to recover equivalence for a large fraction of Maven Central packages rebuilt independently by Google's Assured Open Source (\gaoss) and Oracle's Build-From-Source (\bfs) projects.
The same idea appears, in a more operational form, in \ossr's \emph{stabilisers}: pre-comparison rewrites that normalise away known sources of build variability -- archive entry ordering, embedded timestamps, recorded build paths, and similar -- before two artifacts are compared.
Whether such techniques can recover equivalence at scale in the Python ecosystem remains an open question, and is one we set out to answer in this paper.

The Python ecosystem is a particularly important target.
Python has become the lingua franca of AI and data science, and \pypi is now one of the busiest package registries, distributing packages that are pulled into model training and inference pipelines worldwide.
\pypi has also been the venue of a series of notable supply chain incidents.
The \textit{torchtriton} dependency-confusion attack against \textit{PyTorch nightly}~\cite{torchtriton} demonstrated how the boundary between public \pypi and private package indices can be exploited against AI workloads.
More recently, the compromise of \textit{ultralytics}~\cite{ultralytics2024} -- a widely used computer-vision package -- injected a cryptominer into wheels published to \pypi by hijacking the project's GitHub Actions workflow.


\pypi also has ecosystem-specific properties that make rebuilds harder than in, e.g., Maven.
Sources are typically distributed both as a source distribution (\textit{sdist}) and as one or more pre-built \textit{wheels}. The wheel is what most users actually install, and its relationship to the \textit{sdist} -- let alone to the source repository -- is mediated by a build backend (\textit{setuptools}, \textit{poetry}, \textit{hatch}, \textit{flit}, etc.) that may execute arbitrary Python at build time via \texttt{setup.py} or build hooks.

The \texttt{pyproject.toml} embedded in the published \textit{sdist} is frequently a rewritten version of the file in the source repository, and many packages mix pure Python with C, C++, Rust or CUDA extensions, which require pinned tool-chains to reproduce.
Prior work on Python rebuilds has largely focused on fixing dependency errors~\cite{mukherjee2021fixing} or on \pypi as one of many ecosystems in cross-registry studies~\cite{benedetti2025empirical}. To the best of our knowledge, a focused study using independent rebuild tooling targeted at \pypi is still missing.

In this paper, we study the state of rebuildability of \pypi releases, using \macaron and \ossr as independent rebuild tools on a corpus of popular releases drawn from \pypi.
We answer the following research questions:

\begin{enumerate}
\item [RQ1] For how many releases is a rebuild possible, and what are the most common causes of failure? (Section~\ref{sec:rq1})
\item [RQ2] When a release is rebuildable, is it actually built from the same sources as the original \pypi release? (Section~\ref{sec:rq2})
\item [RQ3] For how many releases is the rebuilt wheel bit-for-bit identical to the \pypi wheel? (Section~\ref{sec:rq3})
\item [RQ4] For how many of the remaining releases can we establish equivalence between the original wheel and the rebuilt wheel? (Section~\ref{sec:rq4})
\end{enumerate}

In addition, Section~\ref{sec:daleq4py} introduces \daleqpy, a tool providing explainable equivalence for Python packages; RQ4 evaluates this tool on a real-world dataset.

The remainder of the paper is structured as follows: Section~\ref{sec:dataset} describes our dataset and pipeline, Sections~\ref{sec:rq1}--\ref{sec:rq4} report on RQ1--RQ4 and introduce \daleqpy, Section~\ref{sec:discussion} discusses security implications, threats to validity, and the availability of our tool and artifacts, Section~\ref{sec:relatedwork} discusses related work, and Section~\ref{sec:conclusion} concludes.

\textbf{Naming Convention Used Throughout This Paper.} We use the term \textit{release} to refer to a wheel, identified by a \textit{package name} \textit{and} a \textit{version}.
\textit{Package} refers to the name part only. Thus, multiple \textit{releases} can share the same \textit{package}, with each release corresponding to a different version.

\section{Methodology}

\subsection{Dataset Acquisition and Construction}
\label{sec:dataset}

The packages in our dataset were selected from a \textit{libraries.io}-derived list of \pypi projects and expanded into concrete releases. The selection was based on \textit{libraries.io}'s \textit{SourceRank} score,
a package-level ranking signal for open-source
packages.\footnote{See also \url{https://libraries.io/pypi/pip/sourcerank}}
 The list was acquired on 7 May 2026. For each selected package, up to five of its latest \pypi releases were retained. The initial set contained 2,500 packages. After applying \pypi name normalisation, which lowercases names and treats hyphens, underscores, and dots as equivalent, this set contains 12,180 releases for 2,447 unique packages.
We then filtered out releases with binary components, as these are usually built for a specific target platform and are therefore difficult to reproduce. 
We refer to releases without such binaries as \textit{pure-Python}. 
Neither rebuild tool used in our study, \macaron or \ossr, builds the non-pure-Python releases.
We identified those releases using Macaron's generated build specification, in particular the \texttt{has\_binaries} field.

\begin{table}[!ht]
\centering
	\begin{tabular}{lllll}
	\hline
		& from libraries.io & deduplicated & pure  &  \\\hline
			packages & 2,500             & 2,447         & 2,130         &  \\
			releases & 12,443            & 12,180        & 10,449        & \\
		\hline
	\end{tabular}
	\caption{Numbers of packages and releases at different dataset-construction stages}
	\label{tab:listinfo}
\end{table}

There are 1,731 releases across 317 packages containing binary components, resulting in 10,449 releases across 2,130 packages in the final dataset. 
Table~\ref{tab:listinfo} summarises those numbers.

\subsection{Rebuilding Releases}
\label{sec:rebuild}

For each release, the pipeline attempted to retrieve the published wheel and metadata from \pypi, then used both \macaron and \ossr to produce independent rebuilds of that release:

\begin{verbatim}
python3 -m pip download --no-deps --only-binary=:all: \
  --dest <pypi_root> <package>==<version>
\end{verbatim}

The later wheel-comparison analyses require a comparable \pypi wheel, i.e., a universal \texttt{*-none-any.whl} artifact for the same release. If such a wheel was not available, the rebuild could still be attempted, but the release was excluded from byte-for-byte and wheel-equivalence comparisons against \pypi.

For \macaron, the pipeline first analysed the \pypi release URL and generated a Docker-based build specification, which was then built to extract the resulting wheel:
\begin{verbatim}
run_macaron.sh -o <macaron_output_dir> analyze \
  -purl pkg:pypi/<package>@<version> --deps-depth=0
\end{verbatim}

For \ossr, the pipeline invoked the local rebuild benchmark on the same release:
\begin{verbatim}
go run ./tools/ctl run-bench attest --local \
  --max-concurrency 1 <benchmark.json>
\end{verbatim}

The rebuild experiments were run on macOS 26.5.1 on an \texttt{arm64} machine with an Apple M3 Pro CPU (12 cores) and 36 GB of memory, using Docker 29.2.1, Python 3.14.5, and Poetry 2.4.1. The \macaron runs used \macaron v0.24.0, corresponding to the commit \texttt{4ddb55e}. The \ossr runs used commit \texttt{b1063e4}. At the time of measurement, the experiment data were $\approx$ 1.8 TB in total, including rebuild artifacts, retrieved wheels, retained source code, etc. The rebuild process consumed approximately 24 days.

\section{RQ1: For How Many Releases Is a Rebuild Possible, and What Are the Most Common Causes of Failure?}
\label{sec:rq1}

First, we measure rebuild success for pure-Python releases. RQ1 counts a release as rebuildable when at least one wheel exists in the corresponding tool output: \texttt{extracted\_wheels/}\footnote{\texttt{artifacts/<package>/<version>/macaron/extracted\_wheels/}} for \macaron and \texttt{wheels/}\footnote{\texttt{artifacts/<package>/<version>/ossr/wheels/}} for \ossr. At this stage, we do not compare the rebuilt wheel with the \pypi wheel. Table~\ref{table:rq1} shows that \macaron rebuilt 7,106 releases (68.0\%), while \ossr rebuilt 5,908 releases (56.5\%).

\begin{table}[H]
	\small
	\setlength{\tabcolsep}{2pt}
	\begin{tabular}{p{1.5cm}p{1.5cm}p{1.2cm}p{1.2cm}p{1cm}}
		\hline
		tool & releases & rebuild success & rebuild failure & success rate \\ \hline
		\macaron & 10,449 & 7,106 & 3,343 & 68.0\% \\
		\ossr & 10,449 & 5,908 & 4,541 & 56.5\% \\
		\hline
	\end{tabular}
	\caption{Rebuild success and failure for pure-Python releases by rebuild tool.}
	\label{table:rq1}
\end{table}

We then classified the unsuccessful rebuilds. For \macaron, this classification uses the reported error type and summary together with the build's \textit{stdout} and \textit{stderr}. For \ossr, we combined \textit{stdout} and \textit{stderr} with the \texttt{Message}, \texttt{Error}, \texttt{Artifact}, and \texttt{Strategy} fields saved in \texttt{firestore.json}\footnote{\texttt{artifacts/<package>/<version>/ossr/output/firestore.json}}. We used an explicit tool-reported error when one was available; when a result contained only a general build failure, we examined the diagnostic messages in the logs. Each failed pure-Python release was assigned to exactly one observed cause, choosing the most specific diagnostic available. The resulting categories describe where the recorded rebuild stopped, which is not necessarily the underlying root cause. Column 3 of Table~\ref{table:failure-taxonomy} gives a representative keyword or phrase for each cause; the classifier uses the corresponding diagnostic patterns rather than only the literal text shown in the table.

For \ossr, the structured metadata records the selected artifact, success flag, tool message, strategy, and timing information. For example, the failed \texttt{litellm@1.83.13} rebuild produced the following metadata:

\begin{lstlisting}[caption={An \ossr \texttt{firestore.json} record snippet.},label={lst:ossr-firestore}]
{
  "Ecosystem": "pypi",
  "Package": "litellm",
  "Version": "1.83.13",
  "Artifact": "litellm-1.83.13-py3-none-any.whl",
  "Success": false,
  "Message": "[INTERNAL] Failed to get upstream generator: unsupported generator: : uv 0.10.7",
  ...
}
\end{lstlisting}

The stages in Table~\ref{table:failure-taxonomy} follow the rebuild process. \ossr first selects the \pypi wheel to use as the upstream artifact. The tools then each identify a repository and commit, infer a build specification, prepare the build environment, invoke the project's build process, and check whether it produces a wheel. The observed causes are grouped by the stage at which the saved evidence shows the rebuild stopped. To clarify how these categories are applied, we next discuss some examples.

For the two tools used to rebuild the same release, failure can occur at different stages. For \texttt{pyspark@4.0.2}, \ossr stopped during \textit{PyPI wheel selection} with \texttt{locating \pypi wheel failed}, so we classified it as \textit{no comparable PyPI wheel}. This selection step is specific to \ossr. \macaron proceeded to the build command, but then reported that \texttt{/src} contained neither \texttt{pyproject.toml} nor \texttt{setup.py}; its failure is consequently classified under \textit{Project build invocation} as \textit{build directory is not a Python project}. The other subcategory under the \texttt{PyPI wheel selection} cause is illustrated by \texttt{pygments@2.18.0}. Here, \ossr looked for \texttt{Pygments\--2.18.0.dist\--info/WHEEL}, whereas the wheel contained \texttt{pyg\-ments\--2.18.0.dist-info/WHEEL}. We classified this as \textit{wheel metadata path mismatch}. Of the 343 releases in this category, 295 differ in case or dash normalisation and 48 have another package-name normalisation difference.

The next stages show similarly direct causes. For \texttt{httpx@1.0.dev1}, \macaron reported no repository information, while \ossr reported no commit; these become \textit{repository not identified} and \textit{commit not identified}, respectively. The \ossr result for \texttt{cairo\-svg@2.7.1} is \textit{repository or commit inaccessible}. Under \textit{Build specification inference}, \macaron failed to generate a build specification for \texttt{azure-identity@1.25.1} and \texttt{ansible@13.6.0}, while \ossr reported an \textit{unsupported build generator} for \texttt{litellm@1.83.13}.

Failures during \textit{Environment preparation} include several distinct dependency problems. For example, \macaron classified \texttt{litellm@\-1.84.0rc1} as \textit{build dependency unavailable} after the build reported that it required \texttt{uv$\_$build==0.10.7}; \ossr reported the same cause for \texttt{mkdocs@1.5.1}, whose build required \texttt{babel}. Missing Python modules appear for \texttt{cloudpickle@3.0.0} and \texttt{mmdet@2.28.2}; dependency resolution fails for releases including \texttt{plover@5.2.0} and \texttt{pysocks@1.7.0}; and \texttt{scienceplots@2.0.1} illustrates a \textit{PEP 517 metadata/requirements failed} result.

\textit{Project build invocation} covers failures after the tool has selected a source tree and attempted to run the project build. The \texttt{pyspark@4.0.2} \macaron case illustrates one subcategory: the selected directory is not buildable as a Python project because it contains neither \texttt{pyproject.toml} nor \texttt{setup.py}. Project configuration or invocation failures include cases such as \texttt{azure-mgmt-power\-biembedded@0.30.0rc5}, where \texttt{setup.cfg} contains an option unsupported by the invoked wheel command, and \texttt{pysnooper@1.2.0}, where requirement parsing fails. Python/runtime incompatibilities include \texttt{ipykernel@6.31.0}, whose build reports \texttt{TypeError: `type' object is not subscriptable}, and platform-specific packages such as \texttt{appnope@0.1.1}. Direct build-command failures are rare; for example, \texttt{azure-cli@2.82.0} reaches the project build in \ossr before the Docker run fails.

\begin{table*}[!ht]
\centering
\footnotesize
\setlength{\tabcolsep}{4pt}
\renewcommand{\arraystretch}{1.08}
\begin{tabular}{p{0.17\textwidth}p{0.25\textwidth}p{0.30\textwidth}rr}
\toprule
rebuild stage & observed cause & keyword/phrase & \macaron & \ossr \\
\midrule
\multicolumn{5}{l}{\textbf{1. PyPI wheel selection}} \\
& no comparable PyPI wheel
& \texttt{locating ... wheel failed}
& N/A & 973 (21.4\%) \\
& wheel metadata path mismatch
& \texttt{Failed to extract upstream ... WHEEL}
& N/A & 343 (7.6\%) \\
\hline
\addlinespace[0.3em]
\multicolumn{5}{l}{\textbf{2. Repository and commit identification}} \\
& repository not identified
& \texttt{Cannot find any repository information}
& 1,400 (41.9\%) & 0 \\
& commit not identified
& \texttt{no git ref}
& 0 & 958 (21.1\%) \\
& repository or commit inaccessible
& \texttt{clone failed ... object not found}
& 1 (0.0\%) & 157 (3.5\%) \\
\hline
\addlinespace[0.3em]
\multicolumn{5}{l}{\textbf{3. Build specification inference}} \\
& build specification generation failed
& \texttt{Error while generating the build}
& 292 (8.7\%) & 0 \\
& unsupported build generator
& \texttt{unsupported generator}
& 0 & 273 (6.0\%) \\
\hline
\addlinespace[0.3em]
\multicolumn{5}{l}{\textbf{4. Environment preparation}} \\
& build dependency unavailable
& \texttt{ERROR Missing dependencies}
& 633 (18.9\%) & 330 (7.3\%) \\
& Python module missing
& \texttt{ModuleNotFoundError}
& 84 (2.5\%) & 46 (1.0\%) \\
& dependency resolution failed
& \texttt{No matching distribution found}
& 7 (0.2\%) & 33 (0.7\%) \\
& PEP 517 metadata/requirements failed
& \texttt{metadata-generation-failed}
& 202 (6.0\%) & 133 (2.9\%) \\
\hline
\addlinespace[0.3em]
\multicolumn{5}{l}{\textbf{5. Project build invocation}} \\
& build directory is not a Python project
& no \texttt{pyproject.toml} or \texttt{setup.py}
& 510 (15.3\%) & 40 (0.9\%) \\
& project configuration or invocation failure
& \texttt{invalid pyproject.toml}
& 98 (2.9\%) & 109 (2.4\%) \\
& Python/runtime incompatibility
& \texttt{SyntaxError}
& 58 (1.7\%) & 3 (0.1\%) \\
& build command failed
& \texttt{build failed: docker run failed}
& 0 & 4 (0.1\%) \\
\hline
\addlinespace[0.3em]
\multicolumn{5}{l}{\textbf{6. Wheel production}} \\
& expected wheel not produced
& \texttt{Failed to stat rebuild artifact}
& 53 (1.6\%) & 1,061 (23.4\%) \\
\hline
\addlinespace[0.3em]
\multicolumn{5}{l}{\textbf{Other}} \\
& insufficient diagnostic information
& \texttt{signal: killed}
& 5 (0.1\%) & 78 (1.7\%) \\

\bottomrule
\end{tabular}
\caption{Causes of rebuild failures. The keyword/phrase column describes a text pattern in \textit{stdout}, \textit{stderr}, and/or \texttt{firestore.json}. The failure rates per tool in the last two columns use each tool's total failures as the denominator: 3,343 for \macaron and 4,541 for \ossr. N/A marks a cause that does not apply to a tool; 0 marks an applicable cause with no observed cases.}
\label{table:failure-taxonomy}
\end{table*}

\textit{Wheel production} happens later in the pipeline. Here, the build invocation has progressed far enough that the tool expects an output wheel, but the expected artifact is absent. For instance, \texttt{azure-common@1.1.24} reached this stage for both tools but did not produce a wheel. For \ossr, \texttt{azure-identity@1.25.1} is another example of this category, while the same release failed earlier for \macaron during build-specification generation. The representative phrase \texttt{Failed to stat rebuild artifact} means that the tool checked the expected artifact path on disk, but no file existed there.

Finally, \textit{Other} is reserved for failures whose saved evidence does not expose a more specific build-stage cause. The remaining \macaron cases are the five \texttt{pex} releases whose logs end in copied-file output rather than a clear terminal error. For \ossr, the remaining cases mostly contain progress output or process termination messages, such as \texttt{signal: killed}, but no project-level diagnostic that would place them in one of the earlier stages.

The distribution of failures differs substantially between the tools. For \macaron, the dominant stage is \textit{Repository and commit identification}, with 1,401 failures (41.9\%). This is followed by \textit{Environment preparation}, with 926 failures (27.7\%), and \textit{Project build invocation}, with 666 failures (19.9\%). These categories capture recurring build-backend, JavaScript tooling, metadata-hook, project-configuration, and Python/runtime diagnostics, leaving only 5 \macaron failures (0.1\%) under \textit{Other}. For \ossr, failures are concentrated at the tool-specific wheel-selection step and at later artifact checks: 1,316 releases (29.0\%) failed during \textit{PyPI wheel selection}, 1,115 (24.6\%) failed during \textit{Repository and commit identification}, and 1,061 (23.4\%) reached \textit{Wheel production} without the expected wheel. The \ossr \textit{Other} row remains small relative to these stages (78 failures, 1.7\%) and represents cases where the retained logs do not expose a project-level cause.

\begin{rqanswer}{Answer to RQ1}
A rebuild was possible for roughly two thirds of the pure-Python releases: \macaron rebuilt 7,106 of 10,449 releases (68.0\%) and \ossr rebuilt 5,908 (56.5\%). The most common cause of failure for \macaron is the inability to identify the source repository or commit (41.9\% of its failures). For \ossr, failures concentrate in the absence of a comparable \pypi wheel (29.0\%), failed repository and commit identification (24.6\%), and builds that complete without producing the expected wheel (23.4\%). Unavailable or unresolvable build dependencies account for a further substantial share of failures for both tools.
\end{rqanswer}

 \section{RQ2: How Many Releases Are Rebuilt From the Original Sources?}
 \label{sec:rq2}
 
A meaningful rebuild should start from the source used to create the published release.
This is the prerequisite in the rebuild equation~\eqref{eq:equiv}.
 RQ2 examines this in two steps. First, we ask whether \macaron and \ossr identify the same repository and commit. We then compare the source selected by each tool with the Python files in the corresponding \pypi wheel.

 Identifying the source of a release is not always straightforward. Both tools draw on package metadata and release-to-commit matching, but a release tag does not necessarily identify the commit used to build the published wheel~\cite{shobe2014mapping,rapaport2026mutating,dietrich2026variability}. When available, \Macaron additionally uses SLSA provenance attestations published through the PyPI Integrity API,\footnote{\url{https://pypi.org/integrity/<package>/<version>/<wheel>/provenance}. The API returns an in-toto provenance statement containing, among other metadata, the source repository URL, commit hash, build workflow (e.g., GitHub Actions), and signing certificates used to attest the build.} which provide build provenance beyond package metadata. We recovered the repository and commit selected by each tool from its recorded build information. \Macaron retains its source checkout, whereas \ossr records the selected repository and commit without retaining the checkout. For \ossr, we therefore cloned the recorded repository and checked out the recorded commit after the rebuild. These checkouts form the source side of our comparisons.

In our artifacts, the source information used for RQ2 is stored with each release. For \macaron, the retained source checkout is stored under \texttt{output/git\_repos/}\footnote{\texttt{artifacts/<package>/<version>/macaron/output/git\_repos/}}. The selected repository and commit are recorded in the generated build specification,\footnote{\texttt{artifacts/<package>/<version>/macaron/output/buildspec/pypi/\allowbreak<package>/macaron.buildspec}} in the \texttt{git\_repo} and \texttt{git\_tag} fields. For \ossr, the selected repository and commit are recorded in \texttt{output/firestore.json}\footnote{\texttt{artifacts/<package>/<version>/ossr/output/firestore.json}}, under the \texttt{repo} and \texttt{ref} fields of the selected \texttt{Strategy}. The corresponding checkout used for comparison is stored under \texttt{source/}.\footnote{\texttt{artifacts/<package>/<version>/ossr/output/git\_repos/source/}} The resolved source root, repository, and commit used in each RQ2 comparison are also recorded.\footnote{\texttt{rq2\_summary/rq2\_detailed\_summary.csv}}

\begin{table}[H]
    \centering
    \begin{tabular}{lcc}
        \hline
        comparison & releases & rate \\ \hline
        both tools rebuildable & 5,066 & -- \\
        repository matching & 4,951 & 97.7\% \\
        commit matching & 4,989 & 98.5\% \\
        repository \& commit matching & 4,880 & 96.3\% \\
        \hline
    \end{tabular}
    \caption{Agreement between \macaron and \ossr source identification for pure-Python releases rebuildable by both tools.}
    \label{table:source-identification-agreement}
\end{table}

We restricted the tool-to-tool comparison to the 5,066 pure-Python releases successfully rebuilt by both tools. A repository and commit were available for every release in this subset. Because GitHub repository URLs are case-insensitive, we convert them to lowercase before comparison. We did not resolve repository redirects. As shown in Table~\ref{table:source-identification-agreement}, the tools agree on the repository for 4,951 releases (97.7\%), on the commit for 4,989 (98.5\%), and on both for 4,880 (96.3\%).

The remaining cases include 115 repository differences and 77 commit differences; 186 releases differ in at least one of the two fields. Some repository differences reflect project moves rather than different source histories. For \texttt{alabaster@0.7.16}, for example, \macaron selected \url{https://github.com/bitprophet/alabaster}, whereas \ossr selected \url{https://github.com/sphinx-doc/alabaster}. GitHub redirects both URLs to the same project history, and both tools selected commit \texttt{f3fdc04}.

Other cases agree on the repository but not the commit. For \texttt{freezegun@1.5.2}, both tools selected \url{https://github.com/spulec/freezegun}. The \texttt{1.5.2} tag resolves to commit \texttt{ba06fa4}, which \ossr selected. \macaron instead selected \texttt{da2885d}. This occurred because the version of \macaron used in our experiments inferred the source commit from the package's SLSA attestation, which pointed to the commit that triggered the release workflow rather than the subsequent commit that updated the project version to \texttt{1.5.2}.\footnote{This issue has since been fixed in \macaron: \url{https://github.com/oracle/macaron/pull/1413}. Our evaluation uses a version of \macaron prior to this fix.} As a result, the source at commit \texttt{da2885d} declares version \texttt{1.5.1}, and the build produced version \texttt{1.5.1}\footnote{\texttt{freezegun-1.5.1-py3-none-any.whl}} instead of the requested version. This example shows that successfully completing a build does not by itself establish that the selected commit corresponds to the intended release. More generally, source identification must account for moved repositories, renamed URLs, mutable tags, and cases in which the inferred release commit does not uniquely identify the source used for publication.

This ambiguity has security implications. A rebuild system that follows a mutable tag, branch, or redirected repository can be misled if that reference changes after publication~\cite{rapaport2026mutating}. 
For instance, GitHub recommends pinning third-party actions to a full commit SHA because tags can be moved or deleted~\cite{github-secure-use}. 
We do not interpret the mismatches observed in our dataset as evidence of malicious activity. Rather, these incidents show why a rebuild must identify its source precisely before it can provide meaningful supply-chain assurance. Provenance and attestations make this choice more auditable~\cite{source-prov}, although the attested commit may describe the state that triggered a release workflow without capturing later transformations made during that workflow. It must therefore be considered together with the recorded build steps and the published wheel.
 
Even when both tools agree, they may not have selected the source used to build the original \pypi release. We therefore perform a second experiment comparing each tool's source checkout and the corresponding \pypi wheel. For each Python file in the wheel, we first look for a repository-relative path with the same suffix and fall back to the filename when no path match exists. We normalise each source (.py) file using an AST-based normalisation. This removes comments and normalises formatting and line endings. This accounts for issues like line-ending normalisation performed by some Git operations.
Finally, we compare SHA-256 hashes of the normalised files. We establish source equivalence when at least one pair of Python files can be compared and all compared pairs have the same normalised hash without a normalisation error. If no comparable pair is found, a selected pair differs, a wheel Python file is unmatched, or normalisation fails, equivalence is not established.

\begin{table}[H]
	\begin{tabular}{lccc}
		\hline
		tool & rebuildable & source equivalent & rate \\ \hline
		\macaron & 7,106 & 5,960 & 83.9\% \\
		\ossr & 5,908 & 5,541 & 93.8\% \\
		\hline
	\end{tabular}
	\caption{Releases with source equivalence established between the original \pypi wheel and the repository identified by each tool (Python sources normalised before comparison).}
	\label{table:rq2}
\end{table}

Table~\ref{table:rq2} shows that we establish source equivalence for 5,960 of the 7,106 releases rebuilt by \macaron (83.9\%) and 5,541 of the 5,908 releases rebuilt by \ossr (93.8\%). The comparison does not establish equivalence for the remaining 1,146 \macaron releases and 367 \ossr releases. This outcome can result from mismatched Python files, a normalisation error, or an unavailable source checkout or \pypi wheel; it does not necessarily mean that the tool selected the wrong repository or commit.

The source root can contain templates, generated files, tests, examples, or Python files that are not present in the wheel, while the build can generate or rewrite Python files. The comparison therefore checks selected source-to-wheel pairs; it does not require every source file to appear unchanged in the wheel.

\begin{rqanswer}{Answer to RQ2}
For the 5,066 releases rebuilt by both tools, \macaron and \ossr agree on the selected repository and commit in 96.3\% of cases. Comparing normalised Python sources against the files in the published wheel, source equivalence with the \pypi release was established for 83.9\% of \macaron rebuilds and 93.8\% of \ossr rebuilds. Most rebuilds therefore do start from the original sources, but precise source identification remains a genuine failure point with security implications, as tags, branches, and repository redirects are mutable.
\end{rqanswer}

\section{RQ3: How Many Releases Can Be Reproduced?}
\label{sec:rq3}

RQ3 asks the stricter question of whether a rebuild reproduces the published wheel byte for byte. For every successfully rebuilt release, we calculate the SHA-256 hash of the rebuilt wheel and compare it with the hash of the corresponding \pypi wheel with the \texttt{*-none-any.whl} suffix. We count the release as reproduced only when the two hashes are identical. A different hash is reported as not reproduced, while releases without a corresponding \texttt{*-none-any.whl} wheel are kept in a separate category. Table~\ref{table:rq3-all} shows that \macaron reproduced 1,092 of its 7,106 rebuilt releases (15.4\%), while \ossr reproduced 1,131 of 5,908 (19.1\%). The hashes differed for 5,544 \macaron releases and 4,777 \ossr releases.

\begin{table}[H]
	\small
	\setlength{\tabcolsep}{2pt}
	\begin{tabular}{lrrrr}
		\hline
			tool & rebuildable & reproduced & not reproduced & no PyPI wheel  \\ \hline
			\macaron & 7,106 & 1,092 & 5,544 & 470  \\
			\ossr & 5,908 & 1,131 & 4,777 & 0 \\
		\hline
	\end{tabular}
			\caption{Reproduced pure-Python releases.}
	\label{table:rq3-all}
\end{table}

\vspace{-0.5cm}

For the remaining 470 \macaron releases, no corresponding \pypi wheel was available in our retained artifacts, so a hash comparison is not possible. The absence of corresponding cases for \ossr should not be interpreted as greater rebuild success. \Ossr requires such a \pypi wheel as an upstream artifact before attempting a rebuild. For all 470 of these releases, \ossr produced no wheel and the failures are reported in RQ1 as \textit{no comparable PyPI wheel}. \macaron, by contrast, rebuilt these releases from source. They therefore remain in its rebuildable population, but are reported separately because there is no published wheel against which to compare them.

\begin{rqanswer}{Answer to RQ3}
Only a minority of rebuilds reproduce the published wheel byte for byte: 1,092 of 7,106 \macaron rebuilds (15.4\%) and 1,131 of 5,908 \ossr rebuilds (19.1\%) have a SHA-256 hash identical to that of the corresponding \pypi wheel. Strict bitwise reproducibility is therefore the exception rather than the rule for \pypi releases.
\end{rqanswer}

\section{\textit{daleq4py} -- Release Equivalence for Python}
\label{sec:daleq4py}

\subsection{Overview}
\label{sec:daleq4py:overview}

The results discussed in Section~\ref{sec:rq3} suggest that the number of releases that can be reproduced is low. 
From a security-analysis point of view, this raises the question of whether those differences indicate a compromised release -- usually the \pypi release, as the rebuild is run in a secure environment -- or result from benign variability and build non-determinism. This mirrors observations made for other languages and software ecosystems~\cite{ren2018automated,shi2021experience,fourne2023s,malka2024reproducibility,dietrich2025levels,dietrich2025daleq,benedetti2025empirical,sharma2025canonicalization}. Rebuilders have reason to investigate such cases: they need to ensure that wheels they distribute to their clients are semantically equivalent to the original releases, and owners of open-source projects may reject maintenance based on alternatively built wheels whose behaviour might differ. Establishing this manually is expensive; tooling is needed to automate the process.

In \cite{dietrich2025levels} a framework was proposed to replace the strict requirement of bitwise equivalence (usually checked by comparing cryptographic hashes) by a weaker notion of equivalence, with \daleq~\cite{dietrich2025daleq} as a proof-of-concept implementation for Java.  In a nutshell, the requirements for a binary equivalence relation $\simeq_{bin}$  (as introduced in equation \ref{eq:equiv}) are: 
(1) Equivalence is a mathematical equivalence relation, i.e., it is reflexive, symmetric and transitive. 
(2) Equivalence under-approximates (undecidable) \textit{behavioural equivalence}. That is, equivalent wheels must have the same behaviour (soundness). 
(3) Equivalence over-approximates bitwise equality. That is, identical wheels are equivalent.
(4) Equivalence is explainable. I.e., provenance is produced to support equivalence decisions made by tools. This could be formal proofs, or tests (witnesses) showing behavioural differences for non-equivalence statements.

Equivalence levels allow the weakening of some of those strict conditions, following the popular idea of code clone levels. This notion embraces existing approaches to normalise (stabilise or canonicalise) wheels before comparing them~\cite{xiong2022towards,schott2024JNorm,sharma2025canonicalization}, as an equivalence relationship can be defined as the function kernel of a normalisation function $norm$ (as long as some provenance is recorded):

\vspace{-0.3cm}
\begin{equation}
	pck_1 \simeq_{bin}pck_2  \iff norm(pck_1)=norm(pck_2)
	\label{eq:norm}
\end{equation} 

But the notion of equivalence is broader. In particular, it can be based on more general functions that produce a canonical representation of a wheel and its semantics. This representation does not necessarily have to be a valid wheel. For instance, in \daleq, the function underpinning equivalence maps Java bytecode to a simplified representation of a relational database that encodes the semantics of this bytecode. Normalisation is implemented by means of a datalog program transforming this database, following best practice in datalog-based static program analysis~\cite{whaley2007bddbddb,bravenboer2009strictly,scholz2016engineering,avgustinov2016ql}. 

The question is how this can be modelled for the pure-Python wheels we want to assess. There are some significant differences: pure-Python wheels do not contain platform-specific compiled extensions, which are excluded by design.


Changes to source code are always suspicious, unless they relate to pure formatting issues. One such issue is line breaks that under certain Git configurations (\texttt{core.autocrlf}) are automatically converted between Unix/Linux/macOS and Windows style when interacting with the repository. Changes to (legal) comments will also change sources.  While this can be easily handled by removing comments and normalising the formatting of sources (e.g., with an AST-based approach), this cannot be expressed (easily) in datalog.  
We found that most of the variability encountered can be traced back to changes in metadata files, such as the \textit{WHEEL} file in the \textit{<package-name>.dist-info} folder. This file uses a simple key-value format, and entries for standard keys such as \textit{generator} often differ across builds.
Another common problem is that any difference in some file is always amplified as it also causes the respective \textit{<package-name>.dist-info/RECORD} files to be different. This file records the sizes and \textit{sha256} hashes of all entries of the respective wheel (except its own).

This means that the datalog representation of a wheel must be fundamentally different from the approach taken by \daleq for Java, where the focus was on comparing compiled classes, i.e., bytecode.  

\begin{figure}[htbp]
	\centering
	\includegraphics[width=0.45\textwidth]{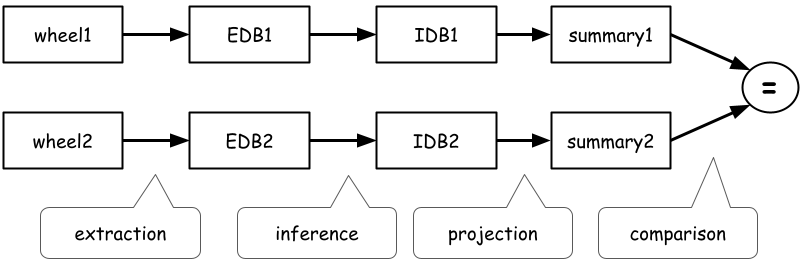}
	\caption{\daleqpy high level design.}
	\label{fig:daleq4py}
\end{figure}

We propose a tool \daleqpy that uses an architecture similar to \daleq, but only uses a single database for the entire wheel. 
The high-level design of \daleqpy is depicted in Figure~\ref{fig:daleq4py}.  The comparison of wheels using \daleqpy can be broken down into four steps:

\textbf{Extraction} For each wheel, a relational database is extracted. Using datalog terminology, we refer to this database as the EDB (extensional database). Each EDB fact has a unique fact id.

\textbf{Inference} Datalog rules are applied to the EDB, creating a second database for each wheel. Using datalog terminology, we refer to this inferred database as the IDB (intensional database). Each IDB fact has a unique fact id; this id is computed when the rules are applied, and encodes its \textit{provenance} -- both the rules and the base facts from which it has been derived.  Souffl\'e~\cite{scholz2016engineering} is  
	used as the datalog engine due to its proven scalability and robustness.
	
\textbf{Projection} The two IDBs computed for two wheels to be compared are then \textit{projected}, i.e., provenance information is removed as it does not describe the program itself, only the process by which this representation was derived. Facts for auxiliary predicates are also removed. This results in simple textual representations (summaries) of the IDBs.

\textbf{Comparison} The summaries are compared for equality (bit-by-bit). If they are equal, then the wheels are deemed to be equivalent.

We discuss each step in more detail in the following sections.

\subsection{EDB Extraction}
\label{sec:daleq4py:edb}

The EDB schema is simple and consists of only two relations: \textit{content} and \textit{package}.
Both are defined in a Souffl\'e component \textit{Edb}. Both have an id slot; the values are created when the EDB is created and are unique.\footnote{The ids are therefore similar to primary keys, but stronger in the sense that values are unique across all relations within the EDB.} The respective definitions are shown in Listing~\ref{lst:edb}.
The \textit{package} relation contains only two facts, with keys \textit{name} and \textit{distinfo}. The respective values are the name of the package, and the name of the \textit{dist-info} folder within it, which contains all package metadata files. This is computed using the simple convention of appending \textit{.dist-info} to the package name. 
The \textit{content} relation represents package content using simple key-value pairs, and the names and locations of files within the package.

\begin{lstlisting}[caption={EDB Predicates},label={lst:edb}]
.comp Edb {
	.decl content(id: symbol, file: symbol, path: symbol, key: symbol, value: symbol)
	.decl package(id: symbol, key: symbol, value: symbol)
}
\end{lstlisting}

Consider the tabular representation of a partial EDB in Table~\ref{tab:edb}, where each row represents a fact instantiating the \textit{content} relation.
The first row (F1) represents the metadata entry \textit{``Generator: setuptools (75.6.0)''} in \textit{absl\_py-2.2.1.dist-info/WHEEL}.  The second entry (F2) (abbreviated) represents an entry in \textit{absl\_py-2.2.1.dist-info/RECORD} recording the hash of \textit{absl\_py-2.2.1.dist-info/WHEEL}. 

\begin{table}[h]
	\centering
	\resizebox{\columnwidth}{!} {
	\begin{tabular}{|l|l|l|l|l|} \hline
		\textbf{id} & \textbf{file} & \textbf{path} & \textbf{key} & \textbf{value} \\ \hline
		F1  & WHEEL   & absl\_py-2.2.1.dist-info       & generator                               & setuptools (75.6.0) \\
		\hline
		F2 & RECORD  & absl\_py-2.2.1.dist-info       & absl\_py-2.2.1.dist-info/WHEEL          & sha256=PZU\ldots \\
		\hline
		F3  & app.py  & absl                          & @sha256                                 & d16269\ldots \\
		\hline
		F4  & app.py  & absl                          & @norm/sha256                            & 5985fb\ldots \\
		\hline
	\end{tabular}
}
	\caption{Sample EDB records instantiating  \textit{content}}
	\label{tab:edb}
\end{table}

This creates a representation of metadata that is easy to reason about. For files not using some key-value format, several pseudo-properties are used.  The names of those synthetic properties always start with ``@''.
The values of the \textit{@sha256} property are the \textit{sha256} hashes of the respective files (example: the F3 record in Table~\ref{tab:edb}). The values of the \textit{@norm/sha256} property are the \textit{sha256} hashes of the respective \textit{normalised} files (example: the F4 record in Table~\ref{tab:edb}).

Normalisation is applied as follows: 

\begin{enumerate}
	\item  For .py files, an AST-based normalisation is applied; in particular, comments are removed, and code formatting and line breaks are normalised to \verb|\n|.
	\item For all other text files identified by an extension whitelist, line breaks are normalised to \verb|\n|.
\end{enumerate}

A particular challenge of this approach is that attackers can spoof the synthetic properties, potentially forcing equivalence between benign and malicious wheels. To prevent this, properties using the synthetic property syntax are renamed and warnings are emitted. 

\subsection{IDB Derivation}
\label{sec:daleq4py:idb}

Given the EDB and a set of datalog rules, an IDB is inferred that is the input into the comparison to check equivalence. 
There is a set of default rules that simply lifts each EDB fact into an equivalent IDB fact.  
The lifting rule for the \textit{content} predicate is shown in Listing~\ref{lst:idb:lift}.  The signatures of the IDB and EDB predicates \textit{content} and \textit{package} are the same; the computed values for the id term in the rule heads will be discussed in Section~\ref{sec:daleq4py:provenance}.  Note that the second premise in the rule body acts as a guard that can be used to remove and/or replace the default lifting behaviour.  

\begin{lstlisting}[caption={Default lifting rule},label={lst:idb:lift}]
	idb.content(cat("R_CONTENT","[",id,"]"),file,path,key,value)  
	    :- 	edb.content(id,file,path,key,value),  
	    		!idb.remove_content(_,file,path,key).
\end{lstlisting}

An example of a custom rule is shown in Listing~\ref{lst:idb:custom1}. 
This rule prevents the \textit{@sha256} property for Python source files from being lifted.
Note that the \textit{@norm/sha256} property is still lifted, and taken into account to establish equivalence.

\begin{lstlisting}[caption={Custom rule to remove @sha256 properties of Python sources (id synthesis omitted)},label={lst:idb:custom1}]
	idb.remove_content(.., name, path, "@sha256") 
		:- 	edb.content(id1,name, path,"@sha256",_),
				match(".*\\.py", name).
\end{lstlisting}

To deal with the redundancies of hashes recorded in \textit{<package-name>.absl\_.dist-info/RECORD}, the normalisation rule in Listing~\ref{lst:idb:removerecord} removes all properties originating from this file.
We consider this to be safe for the following reasons: 
(1) the removal is transparent and recorded, i.e., it is a rule that creates provenance in the IDB (namely, in the exported facts for the \textit{idb.re\-move\_content} predicate);
(2) since we generate EDB facts for \textit{sha256} properties during extraction anyway, the facts extracted from \textit{RE\-CORD} are redundant. 

This decision significantly simplifies the rules, as each rule that normalises some property otherwise needs to remove the respective \textit{RECORD} fact as well. 

\begin{lstlisting}[caption={Custom rule to remove RECORD properties (id synthesis omitted)},label={lst:idb:removerecord}]
	idb.remove_content(.., "RECORD", distinfo, key) 
	   :-  edb.package(id1, "distinfo", distinfo), 
	       edb.content(id2, "RECORD", distinfo, key, _).
\end{lstlisting}

There are several additional rules we implemented. 

\textbf{Build tool identity.}
	The \texttt{Generator} header in the \texttt{WHEEL} file records the tool
	that produced the wheel (e.g., \texttt{setuptools}, \texttt{flit}, \texttt{bdist\_wheel},
	 \texttt{poetry}, \texttt{hatchling}, \texttt{pdm},
	\texttt{maturin}, \texttt{scikit-build}, \texttt{meson}, or \texttt{whey}),
	including its version. If the generator matches a whitelist of known,
	trusted PEP~517 build backends, the header is removed, so that wheels built
	by different (versions of) standard build tools are not considered
	different. 
	
\textbf{Dependency declarations.}
	Raw \texttt{Requires-Dist} headers in \texttt{METADATA} are ignored in
	favour of derived, normalised representations of the declared dependencies,
	so that semantically equivalent requirement specifications (differing,
	e.g., in case or quote type) are considered equal.
	
\textbf{Location of license and author files.}
	\texttt{LICENSE} and \texttt{AUTHORS} files placed directly in the
	\texttt{.dist-info} directory are relocated to the
	\texttt{.dist-info/licenses} subdirectory, reflecting the layout mandated
	by newer packaging conventions (PEP~639), so that wheels differing only in
	where these files are stored are considered equal.
	
\textbf{Metadata format version.}
	The \texttt{Metadata-Version} header is promoted stepwise
	($1.0 \rightarrow 1.1 \rightarrow \ldots \rightarrow 2.5$) to a common
	target version, so that wheels differing only in the metadata format
	version used by their respective build tools are considered equal. When
	promoting from version~2.1 to~2.2, the implicit dynamic headers are made
	explicit: as specified in the core metadata specification, all headers
	present other than \texttt{Name}, \texttt{Version}, and
	\texttt{Metadata-Version} (and, following \texttt{setuptools} behaviour,
	\texttt{Requires-Dist}) are marked with synthesised \texttt{Dynamic:}
	headers.

\textbf{Other text files.}
	For all other text files, identified by a configurable list of file
	extensions, the raw content checksums are replaced by checksums of
	the normalised content, where normalisation is restricted to line-ending
	differences.

\subsection{Provenance}
\label{sec:daleq4py:provenance}

The id expressions in the rule heads encode the derivations. For instance, the expression used in the rule head of the custom rule in Listing~\ref{lst:idb:lift}
is	\textit{cat("R\_CONTENT","[",id,"]")}.
If this rule is applied to a fact with the id \textit{F42}, the derived fact has the id  \textit{R\_CONTENT(F42)}. This encodes that this fact was derived by applying a rule named 
\textit{R\_CONTENT} to a fact with the id \textit{F42}.  This id synthesis is recursive, and can include zero or more rule premises.

There are two limitations: (1) negated premises are ignored, since they do not refer to particular facts, but are quantified over a set of facts; (2) expressions used in rule bodies, such as \textit{match("setup\-tools.*", generator)}, are not tracked. These expressions refer to builtin datalog predicates which are checked on the fly by evaluating ground expressions, and are not associated with facts with a given id.  Users inspecting ids to gauge provenance may inspect the actual rule in order to assess the full derivation. 

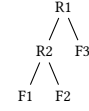
\begin{figure}
	\scriptsize {
	\begin{forest}
		[R1
		[R2
		[F1]
		[F2]
		]
		[F3]
		]
	\end{forest}
}
	\caption{Derivation tree encoded in \textit{R1[R2[F1,F2],F3]} }
	\label{fig:derivation}
\end{figure}

The generated ids are described by a \textit{lark} grammar included in \daleqpy describing a derivation tree. For instance, the expression \textit{R1[R2[F1,F2],F3]} describes the derivation depicted in Figure~\ref{fig:derivation}.

\subsection{Projection}
\label{sec:daleq4py:projection}

IDBs are projected by removing the id terms (first terms) from all derived facts, and removing facts for all auxiliary predicates, such as \textit{remove\_content}. That is, only facts for \textit{content} and \textit{package} are considered.
Facts instantiating auxiliary predicates describe how facts were derived, providing provenance on the normalisation process, but do not describe the normalised structure of the respective wheel itself.  In a final normalisation step, the projected IDB is exported into a plain text file.

\subsection{Comparison}
\label{sec:daleq4py:comparison}

The final step is to compare the projections in text format using a standard diff tool. If those files match, the wheels being compared are considered to be equivalent.

This means that the equivalence relation between wheels is defined as the kernel of the normalisation function \textit{norm}~\cite{dietrich2025levels}:

\vspace{-0.3cm}
\begin{equation}
	pck_1 \simeq pck_2 \iff norm(pck_1) = norm(pck_2)
	\label{eq:daleq4py}
\end{equation}

The normalisation function can be decomposed into functions corresponding to each step described above (EDB extraction, IDB computation, projection), with each step producing artifacts that can be used to verify the process (i.e., the EDB, the IDB including encoded derivations for provenance, and the projection).

Also, unlike other existing normalisation techniques \cite{schott2024JNorm,xiong2022towards}, the normalisation does not produce a valid wheel or code. Its only purpose is to produce a representation of the wheel that is sufficient to establish explainable equivalence.

\section{RQ4: For How Many Releases Can We Establish Equivalence with the Rebuilt Wheel?}
\label{sec:rq4}

As discussed in Section~\ref{sec:rq3}, wheels with different hash values may still be equivalent. In this section, we apply the \daleqpy equivalence method proposed in Section~\ref{sec:daleq4py} to the rebuilt wheels. Our primary analysis compares the output of each rebuild tool with the corresponding published \pypi wheel. We then report a separate comparison between the wheels produced by \macaron and \ossr.



\begin{table}[H]
	\centering
	\small
	\setlength{\tabcolsep}{2.5pt}
	\begin{tabular}{lrrrrr}
		\hline
		tool & source equivalent & equivalent & different  & rate \\ \hline
			\macaron & 5,960 & 3,588 & 2,372  & 60.2\% \\
			\ossr & 5,541  & 4,374 & 1,167 &  78.9\% \\
		\hline
	\end{tabular}
	\caption{\daleqpy equivalence between rebuilt wheels and the corresponding published \pypi wheels, restricted to releases whose source was proven equivalent in RQ2.}
	\label{table:rq4-pypi}
\end{table}

Table~\ref{table:rq4-pypi} reports the primary \daleqpy result using the source-equivalent releases from RQ2 as the denominator. All comparisons in this restricted set completed successfully. \daleqpy established equivalence with the published \pypi wheel for 3,588 of 5,960 \macaron releases (60.2\%) and 4,374 of 5,541 \ossr releases (78.9\%). This is substantially higher than the strict reproducibility measured in RQ3, where only 1,092 \macaron releases and 1,131 \ossr releases had SHA-256 hashes identical to the published wheel.

\begin{rqanswer}{Answer to RQ4}
Among rebuilds whose sources were proven equivalent in RQ2, \daleqpy establishes equivalence with the published \pypi wheel for 3,588 of 5,960 \macaron releases (60.2\%) and 4,374 of 5,541 \ossr releases (78.9\%). Explainable equivalence therefore recovers roughly three to four times as many acceptable rebuilds as the strict bit-for-bit comparison of RQ3.
\end{rqanswer}

\section{Discussion}
\label{sec:discussion}

\subsection{Security Implications}
\label{sec:discussion:security}

Independent rebuilds are ultimately a security measure: a difference between the published and the rebuilt wheel is a signal of a potentially compromised build. The results of RQ3 show that, taken naively, this signal is weak -- with bit-for-bit reproduction rates of 15.4\% (\macaron) and 19.1\% (\ossr), a policy that flags every non-identical rebuild would raise an alert for the vast majority of releases. Each such alert must then be triaged by an engineer.

The results of RQ4 show that most of these alerts are not actionable: a large share of the non-identical rebuilds can be shown to be equivalent to the published wheel. A review process that surfaces them anyway will suffer from \textit{alert fatigue}, a phenomenon well documented for program analysis tools deployed at scale -- when most reported issues are not actionable, engineers lose trust in the tool and start ignoring its reports, burying the true positives among the false alarms~\cite{sadowski2018lessons,distefano2019scaling}. By automatically discharging the provably equivalent cases -- 60.2\% and 78.9\% of source-equivalent rebuilds for \macaron and \ossr, respectively -- \daleqpy reduces the review load to the residual differences that actually merit human attention.

A possible concern is that such automatic suppression introduces false negatives~\cite{ami2024false}, i.e., that a malicious modification is normalised away and a compromised wheel is accepted as equivalent. In our case, this concern is largely unjustified due to the soundness of the analysis: the normalisation rules are designed to only remove information that does not affect the behaviour of the package, so equivalent wheels remain behaviourally equivalent. There is, however, a human element to this claim -- soundness is not machine-checked, but was ensured by carefully reviewing the rules. The explicit, declarative representation of the rules in datalog, together with the provenance recorded for each derivation, is what makes such a review feasible.

There is also a trust dimension. By accepting rebuilds that differ only in benign ways, \daleqpy eliminates the need to trust some upstream tools -- notably the particular build backend used, which plays the role of the compiler in a Ken Thompson-style scenario~\cite{thompson1984reflections}: a wheel built by a different (version of a) trusted backend can still be accepted as equivalent. At the same time, the tool itself adds new dependencies that also need to be trusted; we mitigate this by relying only on established standard library packages.

\subsection{Threats to Validity}
\label{sec:discussion:threats}

\textbf{Provenance gaps.} As discussed in Section~\ref{sec:daleq4py:provenance}, the recorded derivations do not capture negated premises and built-in predicates used in rule bodies. The equivalence decision itself is unaffected -- the rules are still applied correctly -- so this is mainly a user-interface issue: the explanation presented to an engineer is incomplete, and assessing the corresponding derivations requires inspecting the actual rules. How well this form of explainability works for engineers assessing rebuilds in practice needs to be evaluated in future work.

\textbf{Rule incompleteness.} The rule set is incomplete -- there might be additional benign sources of build variability that our rules do not yet cover. The equivalence rates reported in RQ4 are therefore a lower bound: adding further (sound) rules can only increase the number of rebuilds accepted as equivalent, not invalidate the equivalences already established.

\subsection{Tool and Artifact Availability}
\label{sec:discussion:availability}

The \daleqpy tool is available at \url{https://github.com/binaryeq/daleq4py/}.

\section{Related Work}
\label{sec:relatedwork}

\textbf{Reproducible builds.}
The insight that users cannot trust a binary unless it can be independently re-derived from its sources goes back to Thompson's \textit{trusting trust} attack~\cite{thompson1984reflections}, and diverse double-compiling was proposed as an early countermeasure~\cite{wheeler2005countering}.
De Carn\'e de Carnavalet and Mannan demonstrated, in a case study on \textit{TrueCrypt}, how difficult it is to verify builds of security-critical software in practice~\cite{de2014challenges}.
The \textit{Reproducible Builds} project has since developed cross-ecosystem guidance and tooling to eliminate common sources of non-determinism~\cite{reproduciblebuilds,lamb2021reproducible}.
Reproducible builds are also a building block in larger supply chain security architectures: \textit{CHAINIAC} uses collectively verified builds for software-update transparency~\cite{nikitin2017chainiac}, while \textit{in-toto}~\cite{torres2019intoto}, \textit{sigstore}~\cite{newman2022sigstore} and SLSA~\cite{slsa} provide attestation and provenance frameworks against which rebuild results can be recorded; PEP~740 brings such attestations to \pypi~\cite{pep740}.

\textbf{Empirical studies.}
Several studies have measured reproducibility at scale.
Benedetti et al.~studied packages across multiple ecosystems including \pypi, and found reproducibility rates to be generally low; on \pypi, embedded archive timestamps were the dominant cause of failing bit-for-bit comparisons, and reproducibility was largely confined to packages using build backends that fix archive metadata~\cite{benedetti2025empirical}.
Goswami et al.~reported similar issues for npm~\cite{goswami2020investigating}.
For Debian, Bajaj et al.~studied how long unreproducibility issues take to fix and how causes correlate with ecosystem factors~\cite{bajaj2024unreproducible}, and Drexel et al.~report on an independent verifier for Arch Linux~\cite{drexel2024reproducible}.
Malka et al.~rebuilt over 700,000 \textit{nixpkgs} packages and found high but not perfect reproducibility, with embedded build dates accounting for a significant share of failures~\cite{malka2025functional}; they also studied the reproducibility of the build environments themselves~\cite{malka2024reproducibility}.
Xiong et al.~investigated build verifiability for Java-based systems~\cite{xiong2022towards}, and Shi et al.~report industrial experience with verifiable builds of large commercial systems~\cite{shi2021experience}.
Complementing these quantitative studies, Fourn\'e et al.~interviewed Reproducible Builds contributors about the practical and social obstacles involved~\cite{fourne2023s}, and Butler et al.~studied business adoption~\cite{butler2023business}.
A precondition for reproducing a build is being able to build at all; studies on buildability consistently find high failure rates~\cite{tufano2017there,hassan2017automatic}, with Mukherjee et al.~addressing dependency-induced build failures specifically for Python~\cite{mukherjee2021fixing}.

\textbf{Causes of unreproducibility, localisation and repair.}
A line of work analyses \emph{why} builds are not reproducible and attempts to localise or repair the causes.
Ren et al.~proposed \textit{RepLoc} to localise the problematic files causing unreproducible builds~\cite{ren2018automated}, \textit{RepTrace} to identify root causes via causality analysis over system-call traces~\cite{ren2019root}, and later automated patching of build specifications~\cite{ren2022automated}.
Recurring causes identified by these and the aforementioned studies include embedded timestamps and build paths, non-deterministic file orderings, locale and environment leakage, and version drift in the toolchain.
Randrianaina et al.~showed that compile-time options themselves are a source of non-reproducibility in highly-configurable systems~\cite{randrianaina2024options}.
Sharma et al.~categorised causes of unreproducible builds in Java and proposed \textit{Chains-Rebuild} to improve rebuild success rates through canonicalization~\cite{sharma2025canonicalization}.

\textbf{Locating the correct sources.}
Rebuilding a published artifact requires identifying the sources it was built from -- an issue we encountered ourselves and discussed in Section~\ref{sec:rq2}.
Vu et al.~proposed \textit{py2src} to identify the source repository corresponding to a \pypi package~\cite{vu2021py2src}, and \textit{LastPyMile} to detect discrepancies between \pypi artifacts and their source repositories, showing that such differences are common and not necessarily malicious~\cite{vu2021lastpymile}.
Keshani et al.~(\textit{aroma}~\cite{keshani2024aroma}) and Hassanshahi et al.~(\macaron~\cite{hassanshahi2023macaron,hassanshahi2025unlocking}) address the related problem of locating the correct commit for Maven artifacts; we used \macaron in our experiments, which has been shown to outperform \textit{aroma}~\cite{hassanshahi2025unlocking}.
Torres-Arias et al.~showed that even git metadata itself can be tampered with, further motivating independent verification of the mapping between releases and commits~\cite{torres2016omitting}.

\textbf{Normalisation.}
When bit-for-bit equality fails, artifacts can be compared modulo benign variation.
Tools like \textit{diffoscope}~\cite{diffoscope} present in-depth diffs of build outputs, and \textit{strip-nondeterminism}~\cite{stripnondeterminism} removes known non-deterministic bits post-build.
\ossr applies \emph{stabilisers} (also referred to as \emph{sanitisation}) that normalise archive metadata such as entry ordering, timestamps, permissions and compression parameters before comparison~\cite{ossrebuild}.
For Java, Xiong et al.~normalise bytecode to mitigate sources of non-determinism~\cite{xiong2022towards}, and Schott et al.~propose \textit{JNorm}, a bytecode normaliser for similarity analysis~\cite{schott2024JNorm}.
Dietrich et al.~frame such techniques conceptually as \emph{levels of binary equivalence}~\cite{dietrich2025levels} inspired by clone types~\cite{bellon2007comparison}, and propose \daleq~\cite{dietrich2025daleq}, which makes equivalence \emph{explainable} by recording the datalog derivations that justify it -- acknowledging that fully reproducible builds are not always achievable. Our \daleqpy tool transfers this approach to Python.

\textbf{Pinning and provenance.}
Schmid et al.~proposed lock files that pin dependencies and Maven plugins to facilitate high-integrity rebuilds of past releases~\cite{schmid2025maven}; by fully recording the build inputs, such lock files provide provenance that removes much of the ambiguity in build environment reconstruction we observed in RQ1 and RQ2.
We note, however, that if a build plugin is merely a wrapper for a separately installed tool (such as a DSL compiler), pinning the plugin alone is not sufficient.

\section{Conclusion}
\label{sec:conclusion}

In this paper, we studied the state of rebuildability of \pypi releases, using \macaron and \ossr to independently rebuild 10,449 pure-Python releases drawn from popular packages. Our results show that rebuilds succeed for roughly two thirds of releases, with failures dominated by source identification and build environment issues, and that the two tools agree on the selected repository and commit for 96.3\% of commonly rebuilt releases. However, strict bit-for-bit reproduction remains the exception: only 15.4\% of \macaron and 19.1\% of \ossr rebuilds are identical to the published wheel.

To close this gap, we presented \daleqpy, a tool that establishes explainable equivalence between wheels as the kernel of a normalisation function implemented with provenance-preserving datalog rules. Applied to source-equivalent rebuilds, \daleqpy accepts 60.2\% (\macaron) and 78.9\% (\ossr) of rebuilds as equivalent to the published wheel -- roughly three to four times as many as the strict comparison. This substantially reduces the review load caused by benign build variability, allowing engineers to focus on the residual differences that actually merit attention, while the recorded derivations keep each equivalence decision auditable.


\begin{acks}
The first and second authors were supported by a gift from Oracle Inc.
The first author was supported by the New Zealand Ministry of Business, Innovation and Employment (MBIE) Endeavour Fund project RTVU2506~\footnote{\url{https://ssc-fort.github.io/}}.

\end{acks}

\bibliographystyle{plain}
\bibliography{references}

\end{document}